# Variability Modes of Blazars from Intensive Optical Monitoring

G. Tosti[1], S. Ciprini[1], E. Massaro[2], M. Maesano[2], F. Montagni[2], and R. Nesci[2].

[1] *Department of Physics, Perugia University, via A. Pascoli, I-06100, Perugia, Italy*
[2] *Department of Physics, "La Sapienza" University, P.le Aldo Moro 2, I-00185, Roma, Italy*

**Abstract.** In this paper we report the main results of our six year long intensive optical monitoring on blazars ON 231 (W Com), BL Lac, and 3C 273. Blazar intensive optical monitoring is an indispensable tool to understand the correlation with the variability in other bands and to discriminate among the proposed emission models.

## INTRODUCTION

Blazars are Active Galactic Nuclei (AGNs), having a variable non-thermal continuum emission extending from radio to γ-rays. The *Compton Gamma Ray Observatory* (CGRO) discovered that blazars are the most powerful extragalactic γ-ray sources up to GeV energies. Blazar overall Spectral Energy Distributions (SEDs) show a typical double-bump structure. The first one can be peaked either in the IR/optical (low-energy peaked blazars LBL, or "red blazars") or in the UV/X-ray bands (high-energy peaked blazars HBL, or "blue blazars") [1]. The intensity and the polarization of this component exhibit strong modifications, explained in terms of Synchrotron (S) emission from relativistic electrons in the plasma jet. The second spectral component extends from X to γ–rays and it is generally explained by Inverse Compton (IC) scattering of low energy photons, whose origin is still unclear. They can be emitted by the same electron population (Synchrotron Self-Compton, SSC, models) (e.g. [2], [3]) or produced outside the jet (External Compton, EC, models) (e.g. [4], [5], [6]). A detailed study of blazars flux variations may provide considerable information on the emitting region dynamics. The knowledge of the possible variability modes in the optical bands is therefore very useful to understand also the significance of the correlation with the variability observed in other bands. Furthermore, blazar optical monitoring is also an indispensable tool to select the best emission model, because each of them predicts different variations of the high energy flux as response to variations of the low energy part of the SED.

In the optical bands blazars, especially the "red", show large amplitude flares, covering a duration range from hours to years. Unfortunately, the optical data collected in the past, are not suitable to perform a quantitative analysis of the blazar variability, mainly because they were collected either during episodic campaigns on a single object with a typical duration of few weeks, or during long term monitoring programs carried out on many sources but with a low sampling rate on single sources.

To exploit the diagnostic capability of variability studies, in the last six years the Perugia and the Rome groups, together with other teams, have concentrated their observational efforts to obtain well sampled light curves (within the weather conditions limits) for a set of well known blazars. The aim of this work, which is still ongoing, is to obtain good time series from which one can retrieve as much as possible information about the structure of blazar variability on a wide time scales interval.

## THE PERUGIA-ROME OPTICAL MONITORING PROGRAM

In this contribution we will give only a very brief review of the different variability modes observed in three of the best studied blazars in our sample: ON 231 (W Com), BL Lac, and 3C 273. All observations were carried out during the last six years with small telescopes, the Perugia 0.4 m Automatic Imaging Telescope, the 0.5 m and the 0.35 m telescopes operated by the Rome group at the Vallinfreda and Greve astronomical stations. All the telescopes are equipped with CCD cameras and filters for the standard photo-metric band-passes B,V (Johnson) and R, I (Cousins). Photometry is performed using sequences of references stars in the same frames of the sources mainly derived from our own measures but for BL Lacertae we included also some points taken from the literature in order to improve the sampling of the light curve, in particular during the great 1997 outburst.

**ON 231** was intensively monitored since 1994 [7] (Fig.1). The luminosity evolution of the source after 1995 was characterized by a series of main bursts. Rapid variations having time scales of hours/days are always superimposed to the main trends. In April-May 1998 ON 231 had an exceptional outburst and reached the highest recorded level since the beginning of the century [8]. In this occasion *Beppo*SAX observed both the S (at energies less than 3 keV) and IC components up to 100 keV [9]. Moreover, rapid X-ray variability was observed at energies smaller than a few keV. After this episode ON 231 showed a brightness level comparable to the pre-burst phase.

**BL Lac** (Fig.1) had a first flare in the summer 1994 which was followed by a quasi-quiescent phase lasting about three years. In 1997 the source started a new activity period. The great summer 1997 outburst was characterized by variations in the optical flux up to a factor of about 6 in V-band over a time scale of one-two days and by a strong IDV (Intra-Night Variability) activity with a marked colour effect [10], [11]. In 1998 and 1999 the source was still active but it was never observed at the 1997 flux levels. In two *Beppo*SAX X-ray observations (June and December 1999) S and IC components were detected, the former variable on the scale of about 20 minutes [12]. The light curve is characterized by several structured bursts and seems to be produced by an intermittent process.

**3C 273** [13] is a well known $\gamma$-ray sources detected several times by the *CGRO* instruments. The optical light curves is less sampled than those of BL Lac, ON231, and (Fig.1) shows a small amplitude long term variation with an under-sampled flickering component superimposed on it.

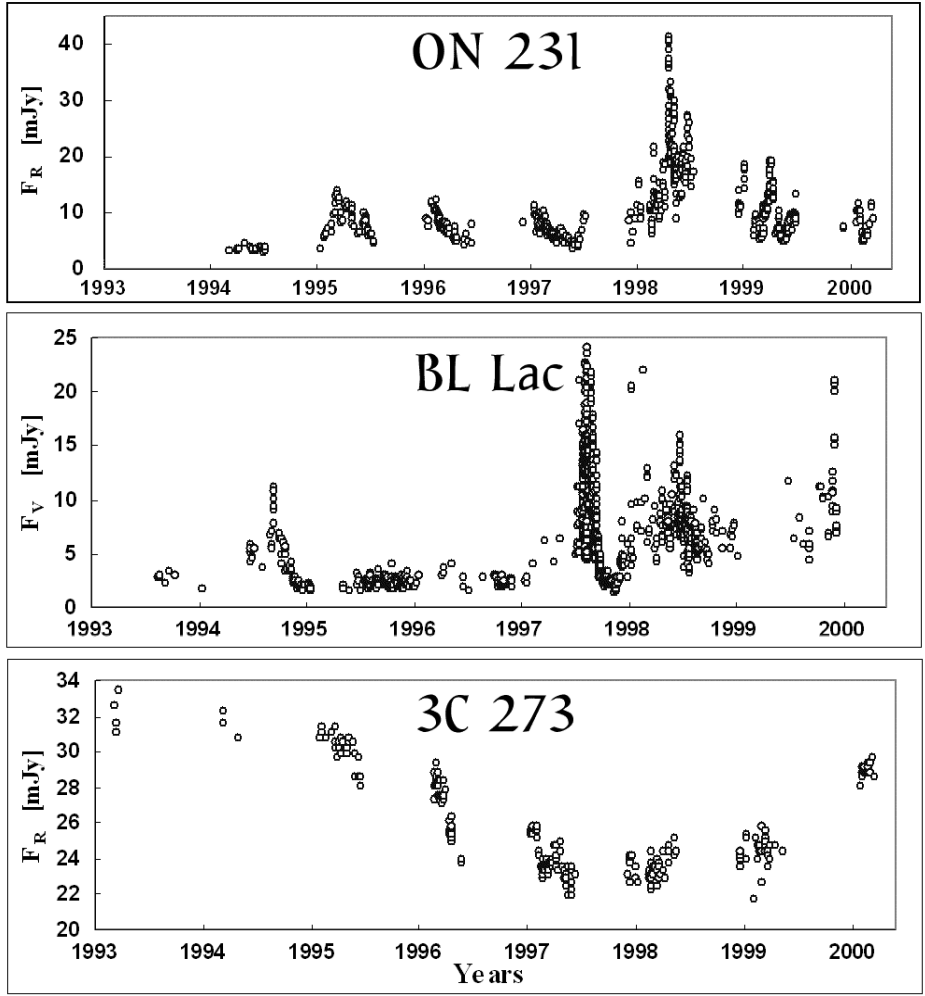

**FIGURE 1.** Optical flux in R-band of ON 231 (W Com), 3C 273, and in V-band of BL Lac, during the period of intensive monitoring observations.

## CONCLUSIONS

The optical light curves described above and based on our monitoring work show different variability modes. We can tentatively identify at least the following three:
- **Intermittent mode**, characterized by a not regular alternation of semi-quiescent and bursting phases; the fine time structure of the bursts shows often a relevant IDV activity (example BL Lac);
- **Quasi-Regular mode**, characterized by the occurrence of long term trend superposed to small scale faster fluctuations (example 3C 273);
- **Mixed mode**, characterized by semi-regular long term trends but showing occasionally large isolated bursts (example ON 231).

We do not know if such different behaviors depend upon the physical conditions in the central object or are simply related either to geometrical or to orientation effects. It is not clear if different variability modes can be present, at various times, in the life of an object suggesting that their occurrence can be related to the evolution of blazars.

Our results demonstrate that a detailed study of the light curves of blazars on time scales of weeks/months/years can be efficiently performed with dedicated small size telescopes. At least for the red blazars, an optimized and nearly continuous optical monitoring is the only way to know the activity status of the sources and to retrieve useful information on their physical dynamics. Also, the rapid availability of information about the luminosity of a source is very important to trigger space based observations and to activate large multifrequency collaborations.

Present and future X and γ-ray space missions carry on board small telescopes for simultaneous optical observations of cosmic sources of high energy radiation. These instruments will provide extremely useful data, but do not reduce the relevance of ground based telescopes because the observations length of the formers is limited by that of the high energy pointing.

We stress that a significant sample of blazars can be observed with small size telescopes (0.35-0.80 cm) equipped with CCD cameras. A world-wide network of several instruments in different countries, is very useful to increase the time coverage and to reduce the number of nights lost for bad meteorological conditions. It is important that the data from all the instruments will be collected by one (or two, for a safer redundancy) group, which will have the task of a fast reduction (using standard methods) and a prompt diffusion. Radio monitoring by Green Bank Interferometer is a good example in this direction. Finally, optical monitoring can profitably help wide field experiments, like either the space mission AGILE [14] or ground based instruments as Argo-YBJ in Tibet [15], able to study the long term variations of several blazars at same time. In particular, we expect that the all these data will be useful to search for luminosity correlation on various time scales.

## ACKNOWLEDGMENTS

The Roma and Perugia groups acknowledge the financial support by the Italian Ministry for University and Research (MURST) under the grant Cofin 98-02-32.